*This manuscript has been authored by UT-Battelle, LLC under Contract No. DE-AC05-00OR22725 with the U.S. Department of Energy. The United States Government retains and the publisher, by accepting the article for publication, acknowledges that the United States Government retains a non-exclusive, paid-up, irrevocable, world-wide license to publish or reproduce the published form of this manuscript, or allow others to do so, for United States Government purposes. The Department of Energy will provide public access to these results of federally sponsored research in accordance with the DOE Public Access Plan(http://energy.gov/downloads/doe-public-access-plan).*

# Stress Induced Charge-Ordering Process in $LiMn_2O_4$

Yan Chen (1), Dunji Yu (1 and 2), Ke An (1)

(1) *Chemical and Engineering Materials Division, Oak Ridge National Laboratory, Oak Ridge, Tennessee 37831, USA.*

(2) *School of Chemical Engineering and Technology, Tianjin University, Tianjin 300072, China.*

In this letter we report the stress induced Mn charge-ordering process in the $LiMn_2O_4$ spinel, evidenced by the lattice strain evolutions due to the Jahn-Teller effects. *In-situ* neutron diffraction, reveals that the initial stage of this process at low stress, indicating the $e_g$ electron localization at the preferential Mn sites during the early phase transition, as an underlying charge ordering mechanism in the charge-frustrated $LiMn_2O_4$. The initial stage of this transition exhibits as a progressive lattice and charge evolution, without showing a first-order behavior.

The spinel-type LiMn$_2$O$_4$ (abbr. LMO) is a basic cathode material for rechargeable lithium-ion batteries [1,2], as well as a frustrated magnetic material [3-5]. The cubic LMO lattice ($Fd\bar{3}m$ space group) has a single crystallographic Mn site, where the Mn ions have an average valence of +3.5. A structural transition accompanying the rearrangement of Mn$^{3+}$ and Mn$^{4+}$ occurs when the temperature is lowered to 280~290 K [5-9], and it consequently leads to lowering the lattice symmetry [10] due to the Jahn-Teller distortion of the MnO$_6$ octahedron with Mn$^{3+}$ that has one electron localized at the e$_g$ band [11,12], as well as abrupt changes of conductivity [13], magnetic property [5] and elastic constants [14]. Although similar to the Verway transition in Fe$_3$O$_4$ [15], the transition of LMO results in a complex charge-frustrated system, in contrast to the alternative charge-ordering configuration of Fe$_3$O$_4$ [16]. Rodríguez-Carvajal *et al* [6] proposed an orthorhombic superlattice (*Fddd* space group) model to fit the average structure of LMO at 230 K, which has five crystallographic Mn sites to accommodate the Mn$^{3+}$ at Mn(1), Mn(2) and Mn(3) sites and Mn$^{4+}$ at Mn(4) and Mn(5) sites (the sites' labels Mn(*i*), *i* = 1~5, are used hereafter). However, this charge-ordering pattern was not considered as the ground state of the electronic system in LMO [6]. The Mn$^{3+}$ and Mn$^{4+}$ have not been distinguishably distributed in the separated Mn sites, and the first-principle study showed that the Mn$^{3+}$ at Mn(2) sites were partially oxidized to satisfy charge neutrality requirements in the *Fddd* model [11]. Single-crystal synchrotron X-ray diffraction at 230K also detected fluctuation of the Mn-O bond-length that indicates a dynamic exchange of oxidation state of Mn between +3 and +4 at Mn(2) sites [12]. Therefore, even at 60 K below 290 K, it is likely that the Mn charge-ordering process in LMO needs a further driving force to reach completion, and that the *Fddd* model still describes one of the intermediate states in the process. Hence, the charge ordering of LMO depends on the thermodynamic state and the process could be resolved progressively as a function of temperature and pressure etc. The question here is how the process starts and evolves. From the previous reports [5,9,10,12], the rapid cubic-to-orthorhombic transition is always difficult to quantitatively resolve during the thermal process, and most think it is a first order transition. Alternatively, stress or pressure is usually a much weaker thermodynamic driving force in comparison to temperature, and could decelerate the transition process and reveal the mechanism of the charge ordering via the lattice response. Previously, *in situ* X-ray diffraction measurements at high pressure up to 20 GPa [17], LMO showed an abrupt transition to the orthorhombic phase at 1.8 GPa and 350 K that is similar as the transition observed under the thermal process, despite the suggestion of formation of a tetragonal phase [18]. To unveil the initial evolving process, low pressure/stress is needed. In this paper, we report that the charge-ordering process can be induced and resolved as a function of continuously applied stress using *in-situ* neutron diffraction [19-21]. The evolving process is followed by monitoring the lattice distortion solely contributed by the Jahn-Teller effects through eliminating the stress induced elastic components. Specifically, the Jahn-Teller effects at the Mn(1), Mn(2) and Mn(3) sites distinguishably result in stretching the octahedron along the three crystallographic axes [001], [100] and [010], respectively [10,12], and therefore, the monitored lattice strain evolution along those axes directly reflect the occurrence of Jahn-Teller distortion, or $e_g$ electron localization, at the particular Mn site. It is found that an initial stage exists in the charge-ordering process in LMO that is a progressive localization of the $e_g$ electrons in the preferential Mn(3) sites, and that the initial stage is triggered at the beginning of loading and leads to a linear stress dependence on the orthorhombic distortion of LMO lattice.

Commercial $LiMn_2O_4$ powders were filled in an aluminum die (10 mm inner diameter) with two stainless steel punches that were initially 16.8 mm apart. The die set with powders was mounted in the VULCAN diffractometer [22-24] at the Spallation Neutron Source (SNS), Oak Ridge National Laboratory (ORNL), and the *in-situ* neutron diffraction experiment was conducted (FIG. 1). Load was applied via the punches with a rate of 50 N/min (~0.64 MPa/min) until the stress reaches -300 MPa (the minus sign stands for compression) and then unloaded with the same rate. The incident neutrons were focused at the center of the powders in the die set, which did not move due to the simultaneous compression displacement from both sides. The measureable volume was defined to about $4\times6\times5$ mm$^3$ by the incident slits and radial collimators. During the mechanical test, the two detector banks continuously recorded the diffracted neutrons with the scattering vector parallel to the longitudinal ($Q_1$) and transverse ($Q_2$) directions, respectively. The diffraction patterns were averaged with a 20 min interval using the VDRIVE software [25]. The lattice parameters were extracted from Rietveld refinement of whole patterns using GSAS and EXPGUI software [26,27]. The measurement at unloading was used to decouple the elastic strains from the total lattice strain, and the datasets at $Q_1$ and $Q_2$ directions were used to verify the elimination of the elastic components.

The LMO lattice response upon the applied stress is revealed as an orthorhombic distortion. Before loading, the LMO lattice is close cubic, but with a measurable orthorhombic distortion. Overall, the orthorhombic model (*Fddd*) is proved to provide better fitting to the neutron diffraction pattern than the cubic model ($Fd\bar{3}m$) (FIG. S1 and Table S1 [28]). FIG. 2(a) shows the comparison of the fitting around the (400)$_c$ peak (the subscript "c" stands for using the pseudo-cubic coordinates). The calculated lattice parameters of this pristine phase are $a$ = 24.720(3) Å, $b$ = 24.762(3) Å and $c$ = 8.2171(7) Å, respectively. The distortion became more significant upon the increase of the applied stress, and the orthorhombic model fits better at the maximum stress of 300 MPa (FIG. 2(a)). Rietveld refinement of the neutron diffraction pattern does, however, not support the fraction changes between two phases. Rather, the phase transition is attributed to the increase of the lattice distortion in the spinel. The occurrence of an orthorhombic distortion is further indicated by the anomalous peak broadening behaviors at different crystal plane directions: less broadening along the body diagonal in the pseudo-cubic lattice, i.e. $(222)_c$, and more along the lattice edge, i.e. $(400)_c$ (FIG. S2 [28]). In the absence of a phase transition, the bare thermal expansion or elastic strain response is unlikely to contribute to this anisotropic broadening. FIG. 2(b) shows the superlattice of the orthorhombic LMO [6]. Although the fitting agrees with the orthorhombic model, thermodynamically, the stress of 300 MPa is still relatively low and may not establish the local atomic displacement and octahedral distortion in a long-range periodicity, without showing the appearance of the characteristic peaks of the *Fddd* superlattice [6,7] in the diffraction patterns (FIG. S1 [28]). Therefore, the loading test scopes at the very beginning of the orthorhombic distortion from the cubic lattice, and it will be at the "initial stage", as named, of the charge ordering. At this stage, it is possible that the $e_g$ electron localization has a preference among Mn(1), Mn(2) and Mn(3) sites to be partially ordered since the three $Mn^{3+}$ sites have different symmetries (FIG. 2(b)). Due to the different Jahn-Teller distortions at Mn(1), Mn(2) and Mn(3) sites [10,12], as indicated by the arrows in FIG. 2(b), the following quantitative analysis of the lattice strains will reveal the details of the charge-ordering process in the initial stage.

The relative lattice change or lattice strain is quantified as the, $\varepsilon = (L - L_0) / L_0$ along the three principal (i.e. crystallographic) axes, as a function of the applied stress (FIG. 3), where the reference $L_0$ is under 0 stress. The orthorhombic lattice distortion due to the charge ordering contributes to an abnormal appearance from a pure elastic or reversible strain-stress relation. In the longitudinal direction under the compression, along the *a*- and *c*-axes the lattice exhibits a response of compressive strain; however, a tensile strain is observed along the *b*-axis even from the beginning (FIG. 3), contradicting the normal elastic behaviors in the metal or ceramic materials [21,29]. Similar abnormal trends of the lattice strains are also observed in the transverse direction (FIG. S3 [28]), without showing a normal Poison's effect. Therefore, it is apparent that the phase transition strain due to the charge ordering ($\varepsilon^{PT}$) contributed to the measured lattice strains, resulting in the abnormal stress lattice-strain behavior. The $\varepsilon^{PT}$ component is the average effect of the Jahn-Teller distortions of the $MnO_6$ octahedra where an $e_g$ electron is localized during this phase transition. The transition here is considered irreversible after unloading from the peak stress 300 MPa, because the electron mobility carries an activation energy [30], and a certain driving force, e.g. overheating, is required to trigger the phase transition back to the cubic lattice [13,17]. Therefore, the $\varepsilon^{PT}$ component is differentiated from the reversible elastic strain, and the pure effects of Jahn-Teller distortions can be extracted to reveal the charge-ordering process.

The unloading curve, that is the recovery of the elastic deformation, is employed as the reference to subtract the contribution of elastic strain. As one can see in the unloading curves, the slopes indicate a tendency of expansion in all the three principal axes at the longitudinal directions in response to the reduction of the applied compressive stress (FIG. 3) while a tendency of shrinkage at the transverse direction (FIG. S3 [28]). Thus, the unloading process is confirmed to be consistent with a normal elastic behavior having an anisotropic dependence upon the loading direction. By subtracting these elastic components and estimating the residual strains (see Supplementary Material [28] for details), the $\varepsilon^{PT}$ along the three principal axes are extracted and plotted as functions of the applied stress in FIG. 4.

Although the strain-stress curves in FIG. 3 are rather complex, the $\varepsilon^{PT}$ components in FIG. 4 turn out to be consistent in both the longitudinal and the transverse directions and have a simple and linear dependence upon the stress. The isotropic behavior confirms that the anisotropic elastic strain component has been excluded and that the occurrence of the phase transition is the major reason for the $\varepsilon^{PT}$. As a result, the pure orthorhombic distortion is concluded to exhibit a tensile strain and a compressive strain along the *b*- and *c*-axes, respectively, while the *a*-axis exhibits only slight compression. Accordingly, the lattice volume of LMO decreases slightly in order to accommodate the applied compressive stress in the container. The tendency of the lattice distortion agrees well with the Jahn-Teller effect on the Mn(3) site that elongates Mn-O bonds along the [010] direction but shortens those along the [001] direction [10,12], as illustrated in FIG. 2(b). This indicates that the Mn(3) sites are preferred for the localization of the $e_g$ electron in the initial stage of the charge-ordering process. The evolution of $\varepsilon^{PT}$ represents the progress of arranging $Mn^{3+}$ at the Mn(3) sites.

In view of the lattice distortion, the charge ordering is induced at the beginning of loading and at the limit of cubic phase of LMO. In this initial stage, the applied stress continuously induces the charge-ordering process; neither an onset of the applied stress nor a discontinuity of the strain is observed, in contrast to the first-order transition appearance observed by DSC [10] that has a distinct phase transition temperature. The initial stage is difficult to observe without a significant indication of the structure transition; however, it is triggered earlier than generally thought. The initial stage that is captured here under loading also exists in the temperature-induced phase transition and charge ordering of LMO. A pair distribution function study [31] revealed the occurrence of local Jahn-Teller distortion at 300 K, higher than the generally accepted transition temperature, in which the $Mn(3)O_6$ octahedra were found to have the largest

Jahn-Teller distortion at this stage, in agreement with our conclusion. The initial stage of the charge-ordering process also coincides with the fact that the short-range magnetic correlation starts building up at room temperature, which is evident from the measured Weiss temperature of 300 K [5], and above the accepted temperature for the structural transition at 280~290 K.

The initial stage is a continuous process such that the Jahn-Teller effects of the Mn(3) sites is monotonically increasing as the compressive stress increases. It is unlikely the "fraction" changes between two ground states in this frustrated system, because it is still not the ground state of the electronic system in the orthorhombic LMO. Since the low level of stress is able to continuously align the $e_g$ electrons at the Mn(3) sites, this suggests that there are multiple excited states without distinct energy difference. The applied stress may progressively drive the system across those states, accompanying with the gradual orthorhombic distortion. Although our loading experiment has not reached the end of the initial stage, it is reasonable to predict that the $e_g$ electron localization at Mn(1) and Mn(2) sites probably occurs with higher driving force at the later stages of the phase transition. It may be noted that effects of Jahn-Teller distortion on Mn(1)$O_6$ and Mn(2)$O_6$ bring elongation along [001] and [100] directions, respectively (FIG. 2(b)) [10,12], which show a different tendency of $\varepsilon^{PT}$ from that in FIG. 4. Including those effects may contribute to the nonlinear lattice parameter changes and anisotropic shrinkage/expansion along the different principal axes, which were observed in the experiments with applying higher pressure [17] or lower temperature [10] that has passed the "first-order" transition onset. Therefore, the charge-ordering process includes multiple stages, and the $e_g$ electron is preferentially localized at the particular Mn site at each stage. At the beginning of the cubic-to-orthorhombic phase transition, although some key stages are rapidly passed during the thermal process showing the appearance of a "first-order" phase transition, the initial stage is revealed with the stress loading. Since the electron configuration of Mn ions in the LMO can be gradually altered under the stress, we predict that the electrical and magnetic properties of LMO will exhibit successive change accordingly in the initial stage, and it will be investigated in the future work.

In summary, *in-situ* neutron diffraction captures the charge-ordering process through the progressive orthorhombic distortion in $LiMn_2O_4$ under loading. An initial stage is revealed during which the stress continuously induces the localization of $e_g$ electrons preferentially at the Mn(3) sites, which is demonstrated by the pure lattice distortion due to the Jahn-Teller effects. The charge-ordering process is triggered at a low stress level and exhibits a linear dependence on the stress in the initial stage. The proposed mechanism provides a new understanding of the charge-ordering process in spinel-type frustrated systems. The stress induced structure evolution also provides important considerations of the physical compatibility and the electrical property change for the material processing and application in batteries.

This work was supported by Division of Materials Sciences and Engineering, Office of Basic Energy Sciences (BES), U.S. Department of Energy (DOE). Neutron work at ORNL's SNS was sponsored by the Scientific User Facilities Division, BES, DOE. The authors thank Mr. M. J. Frost and Mr. H. D. Skorpenske from SNS for their technical support of the neutron experiments. The authors also thank Mrs. Gumin Zhu for the technical support.

**Figures**

FIG. 1. A schematic illustration of the *in-situ* loading setup and the loading curve. The non-linear loading curve with a large displacement includes the process of powder close packing. The nearly linear unloading curve presents the elastic deformation reversion, but the displacement is not fully recovered.

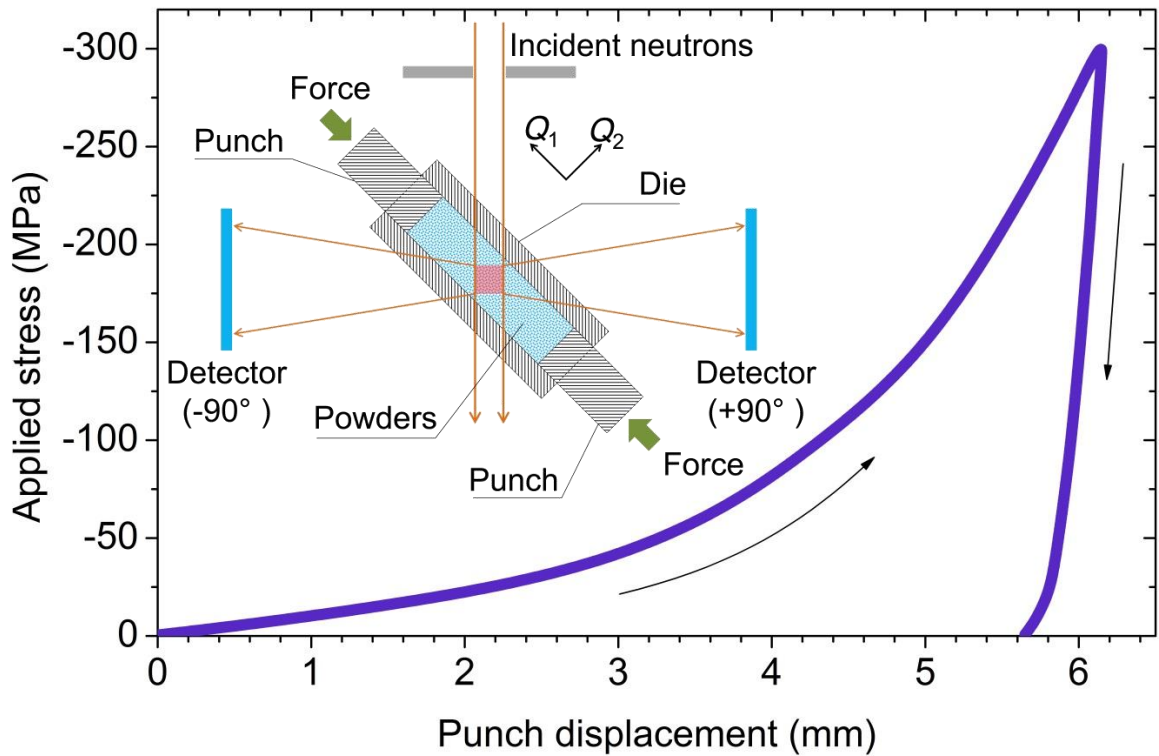

FIG. 2. (a) Illustrations of {400}$_c$ peak fitting by applying the cubic structure model ($Fd\bar{3}m$) and the orthorhombic structure model (*Fddd*) respectively in the neutron diffraction patterns. The purple curves are the fitting residuals. The peak width is best using the *Fddd* over the cubic model confirming the orthorhombic structure of LMO during loading. The Al peaks are due to the Al die. (b) The superlattice and MnO$_6$ octahedra arrangement in the orthorhombic LMO. The Li and O atoms are omitted for simplicity. The grey arrows represent the distortion tendency of the Jahn-Teller effect at Mn(1), Mn(2) and Mn(3) sites respectively.

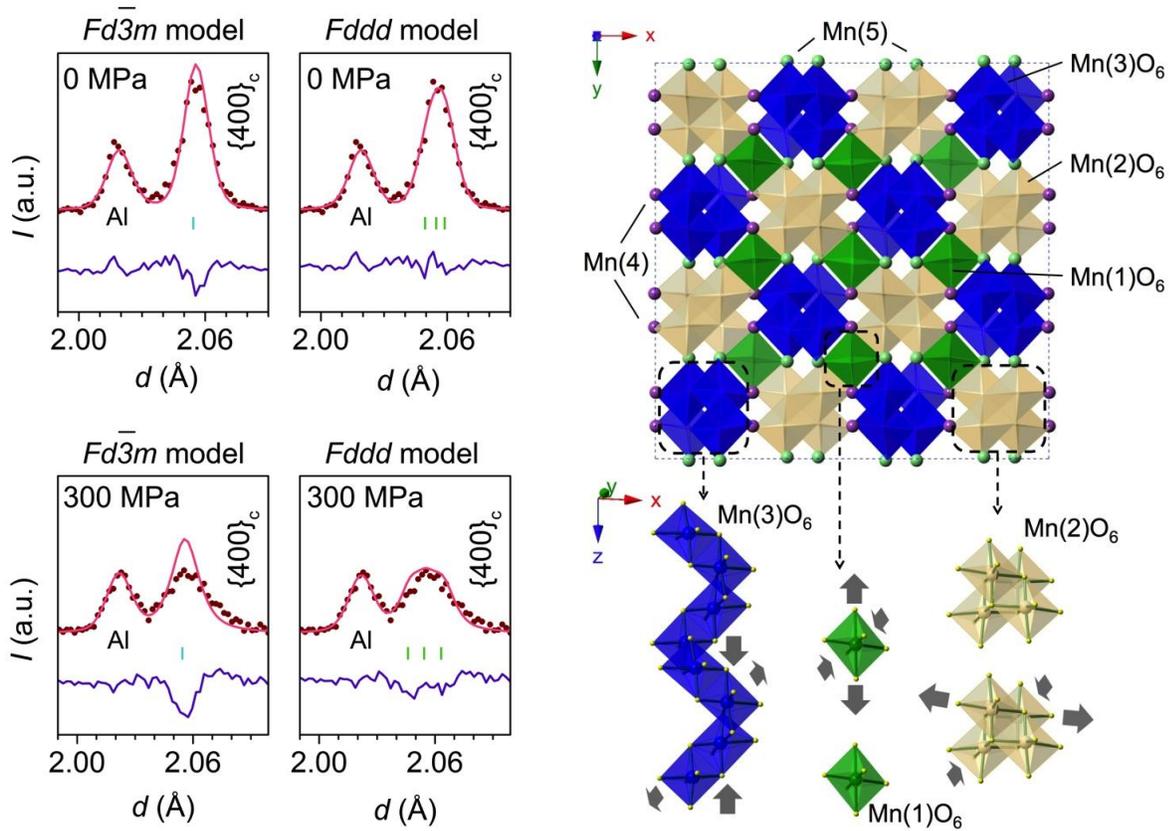

FIG. 3. The lattice strains $\varepsilon$ versus applied stress $\sigma$ curves ~~at~~ along the longitudinal direction ($Q_1$). The loading leads to compressive strains along the $a$- and $c$-axes while a tensile strain along the $b$-axis. The lighter symbols present the data during unloading.

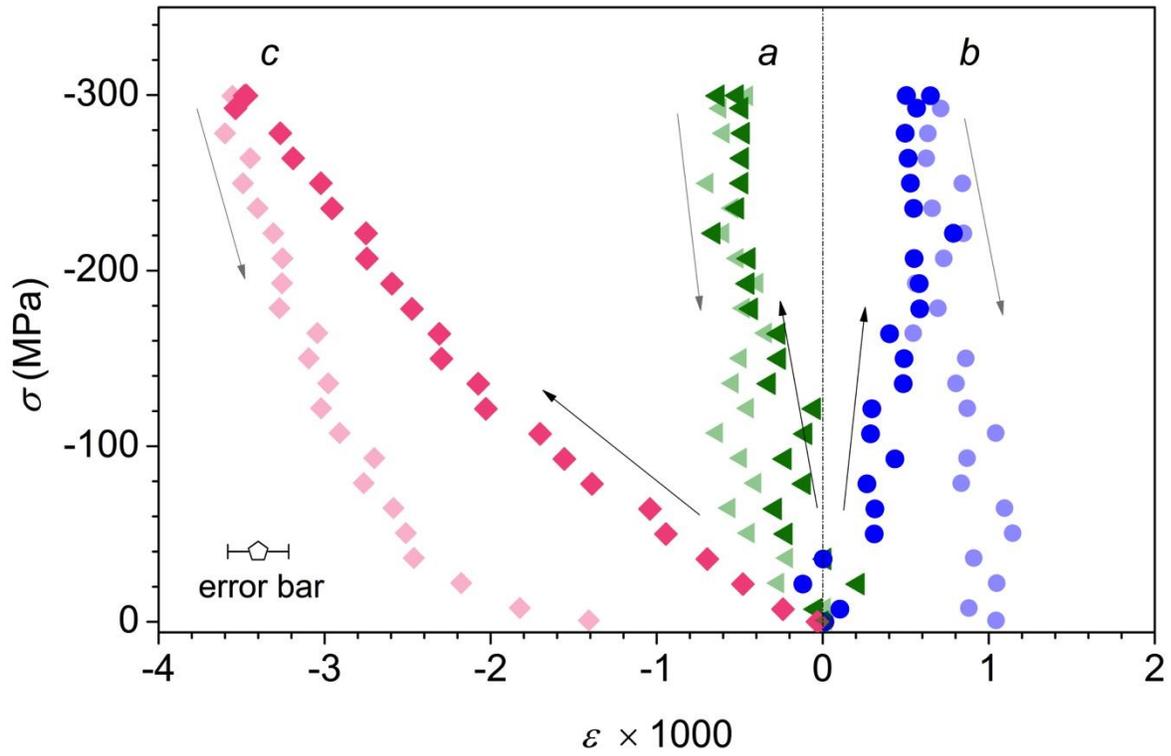

FIG. 4. The phase transition strains $\varepsilon^{PT}$ in the orthorhombic lattice exhibit nearly linear dependence upon the applied stress $\sigma$, without anisotropic dependence upon the loading direction. The volume change rate is calculated through the linear fitting of the strains. The solid and open symbols present the data of longitudinal direction ($Q_1$) and transverse direction ($Q_2$), respectively.

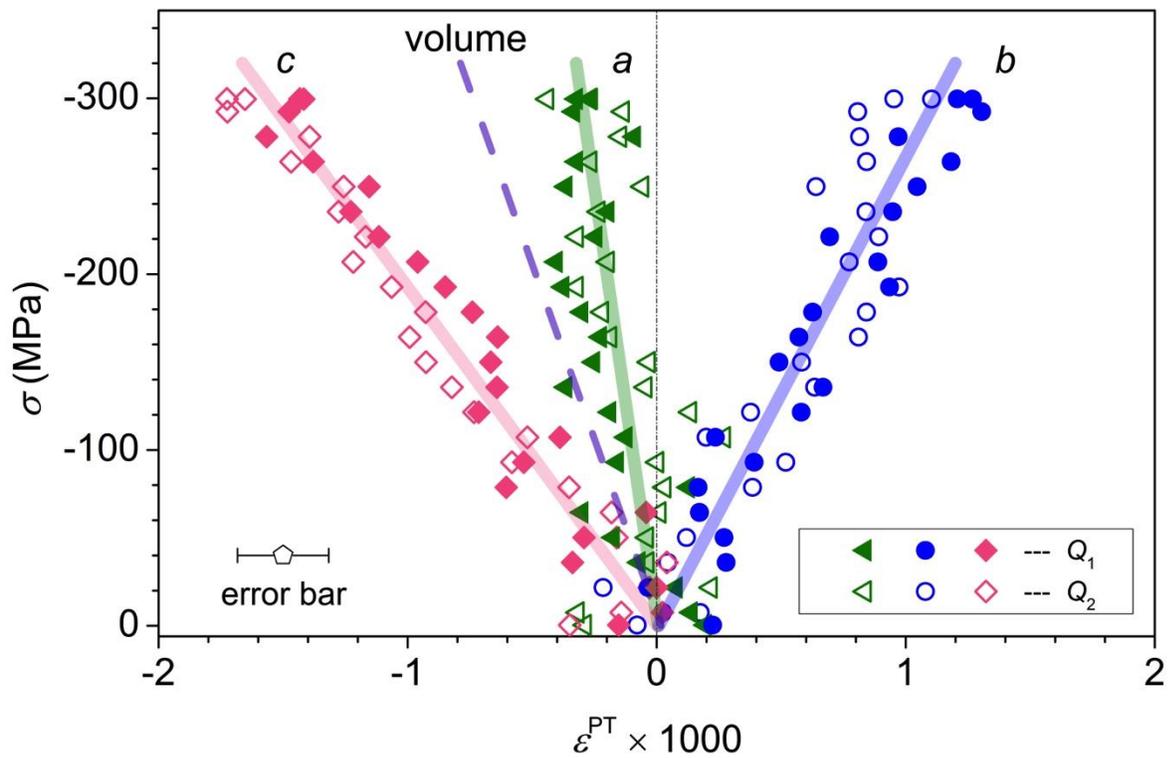

**Supplemental material**

**Stress Induced Charge-Ordering Process in LiMn$_2$O$_4$**

Yan Chen (1), Dunji Yu (1 and 2), Ke An (1)

(1) *Chemical and Engineering Materials Division, Oak Ridge National Laboratory, Oak Ridge, Tennessee 37831, USA.*

(1) *School of Chemical Engineering and Technology, Tianjin University, Tianjin 300072, China.*

FIG. S1. Neutron diffraction and Rietveld refinement using the orthorhombic structure model (*Fddd*). The insets show the details of (400)$_c$ peak fitting, in comparison with the fitting using the cubic structure model (*Fd$\bar{3}$m*). The Al peaks are due to the Al die.

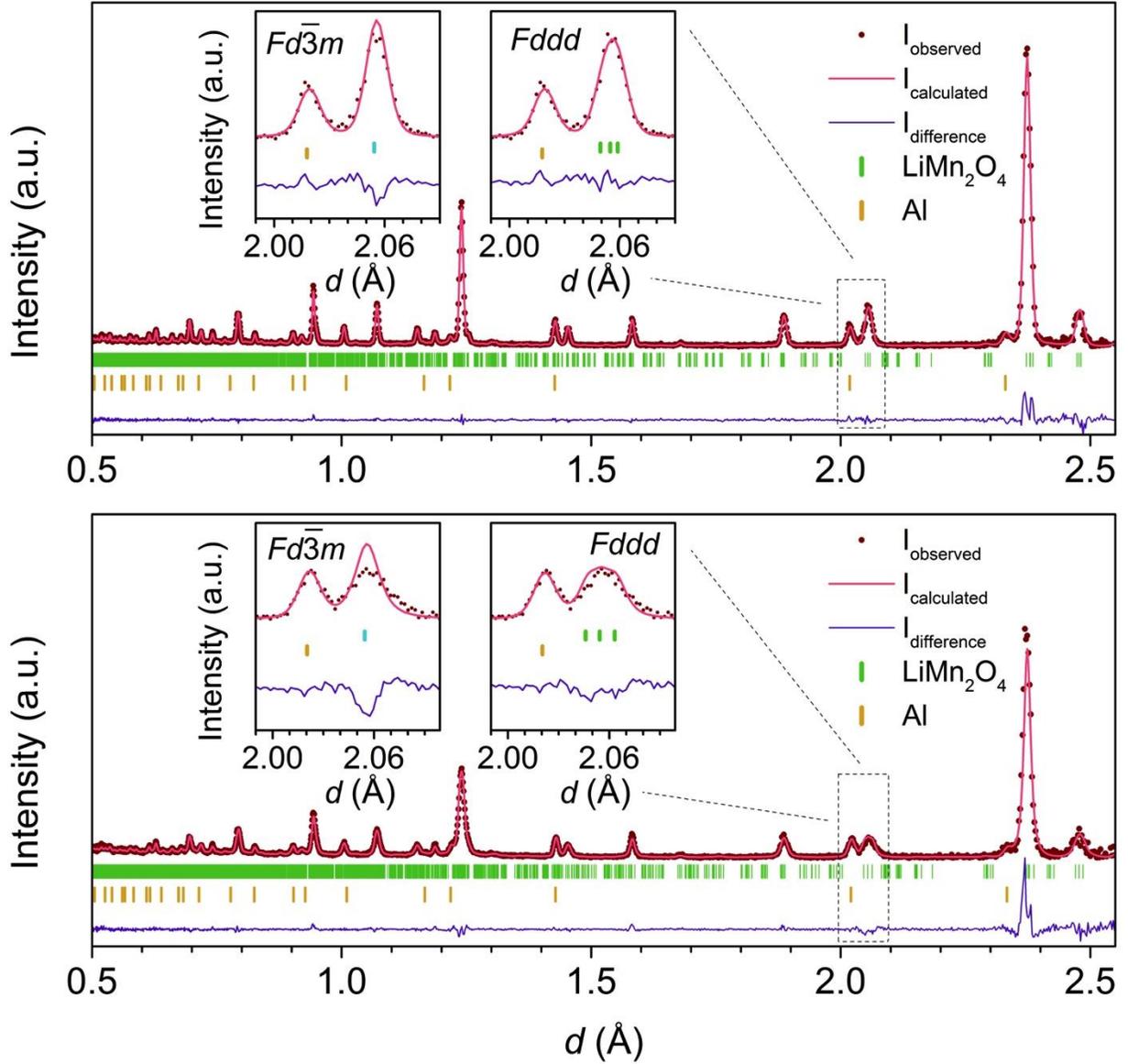

Table S1 Fitting parameters of Rietveld refinement of neutron diffraction patterns in FIG. S1.

| Applied stress | Structure model | $\chi^2$ | %wRp | %Rp |
|---|---|---|---|---|
| 0 MPa | Fd-3m | 1.480 | 8.47 | 6.27 |
| | Fddd | 1.335 | 8.04 | 5.47 |
| 300 MPa | Fd-3m | 2.314 | 10.12 | 7.61 |
| | Fddd | 1.815 | 8.96 | 6.77 |

FIG. S2. The full width at half maximum (FWHM) changes of selected diffraction peaks as a function of the applied stress during loading. A great discrepancy is observed between different diffraction peaks. {$h00$} reflections show the largest peak broadening while the {$hhh$} shows the least, almost maintaining zero. Such discrepancy is caused by more than just the defects and/or microstrains due to loading. It is consistent with the effect of an orthorhombic distortion from a cubic lattice with nearly constant volume. The distortion starts at the beginning of loading.

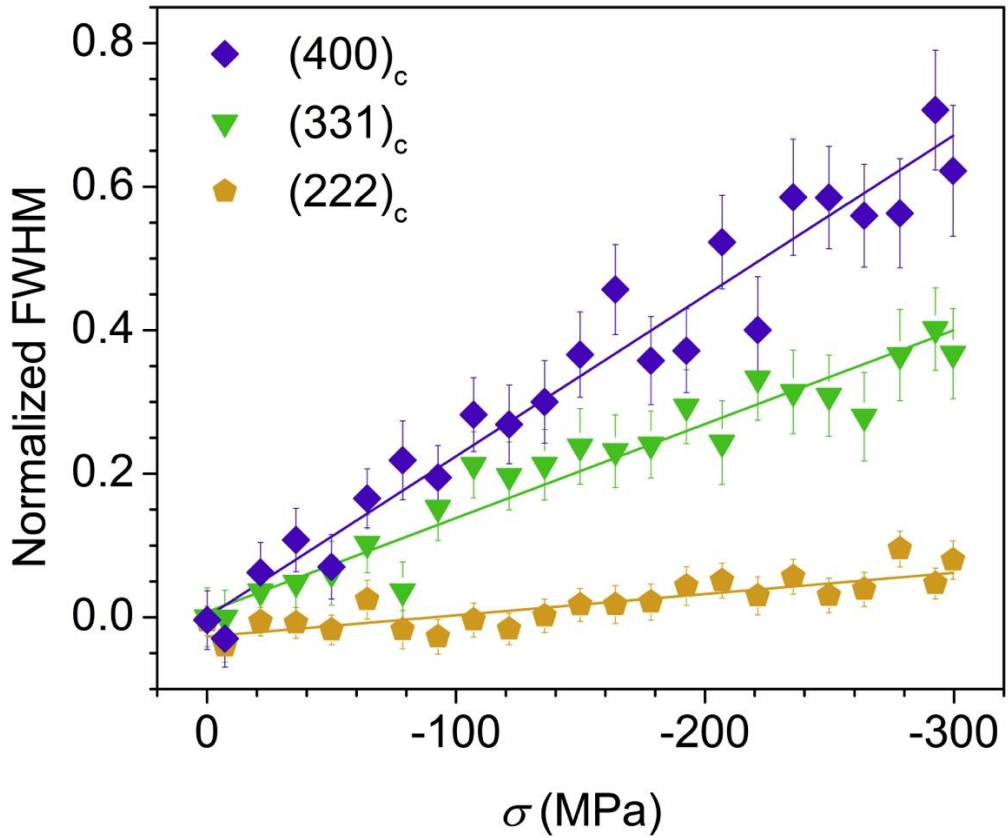

FIG. 3. The lattice strains ε versus applied stress σ curves at along the transverse direction ($Q_2$). The loading leads to compressive strains along the *a*- and *c*-axes while a tensile strain along the *b*-axis. The lighter symbols present the data on unloading.

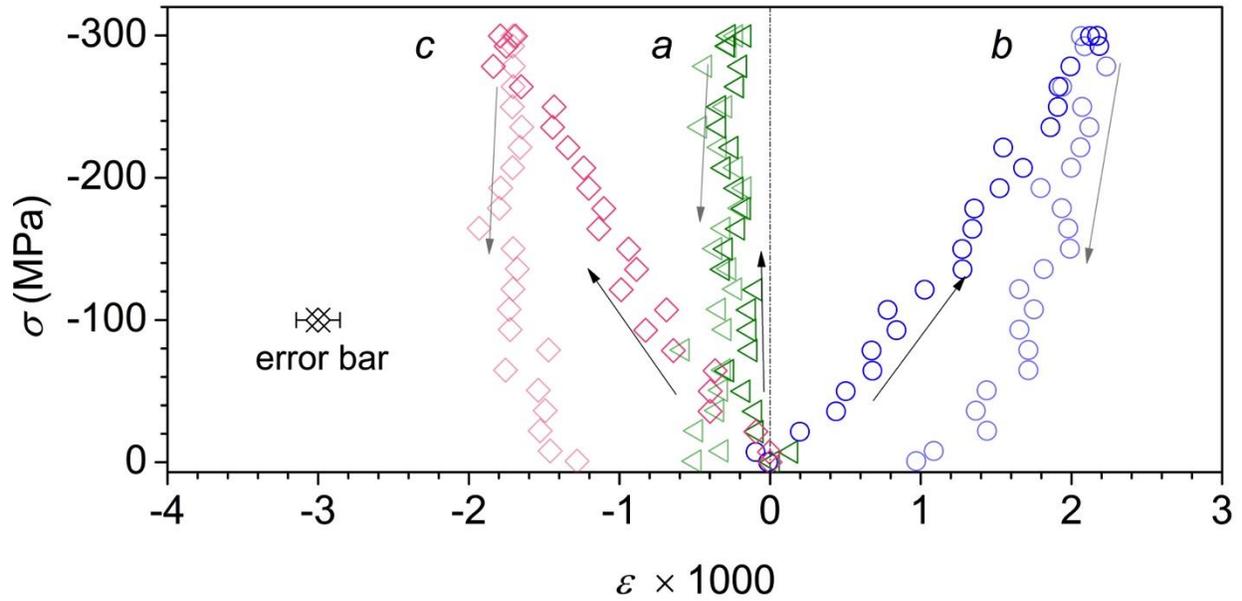

The lattice strains on loading ($\varepsilon_L$) includes the elastic strain ($\varepsilon^{el}$) and the phase transition strain ($\varepsilon^{PT}$) in response to the applied stress. That is

$$\varepsilon_L = \varepsilon^{el} + \varepsilon^{PT}$$

During unloading, the elastic component is recovered while the maximum of the phase transition strain persists that is revealed as the residual strain $\varepsilon_r$. The lattice strains on unloading ($\varepsilon_{UL}$) is expressed as

$$\varepsilon_{UL} = \varepsilon^{el} + \varepsilon_r$$

Therefore, the phase transition strain can be obtained by

$$\varepsilon^{PT} = \varepsilon_L - \varepsilon_{UL} + \varepsilon_r$$

The $\varepsilon_L - \varepsilon_{UL}$ versus $\sigma$ curves are plotted in FIG. S4, which show linear dependence. The residual strains $\varepsilon_r$ are estimated by the intercepts of the linear fitting.

FIG. S4. $\varepsilon_L - \varepsilon_{UL}$ versus applied stress $\sigma$ along the longitudinal direction (solid symbols) and the transverse direction (open symbols). It is found that the data from the two directions are consistent with each other, which indicates isotropic behavior without a significant dependence on the loading direction. The dependence on the stress level is almost linear throughout the applied stress range. The intercept of the fitting line gives $-\varepsilon_r$.

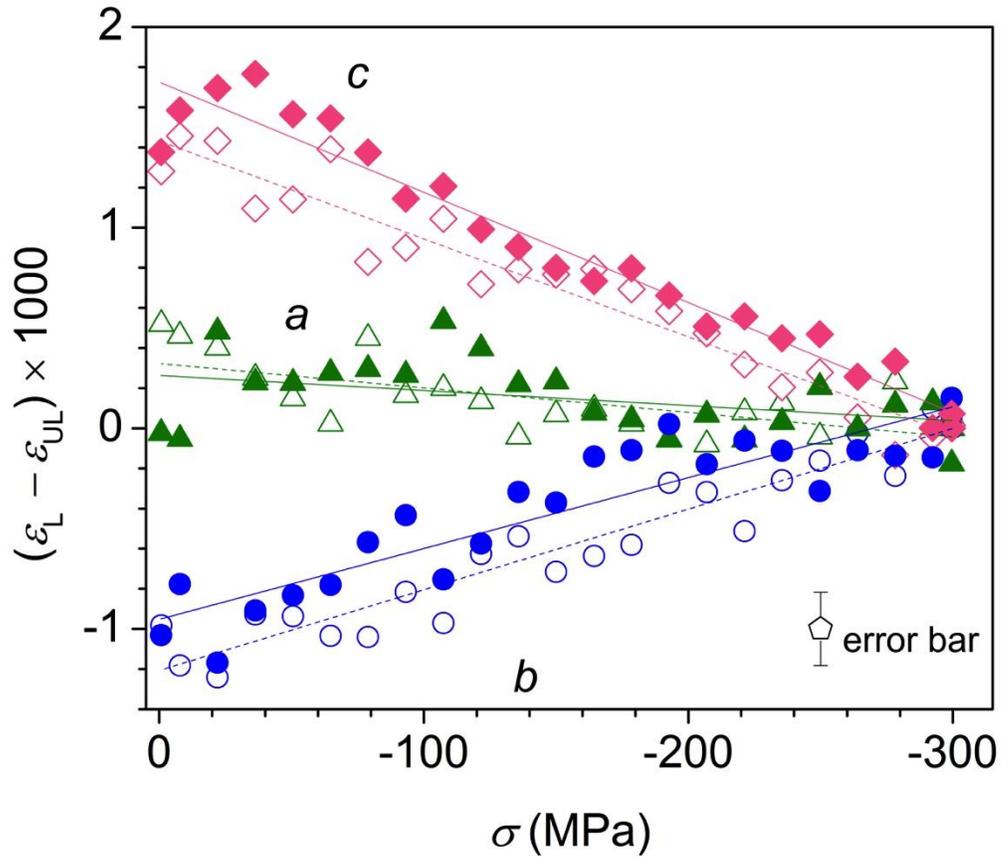